# Experimental study of machining system: dynamic characterization

Miron Zapciu[a], Olivier Cahuc[b], Claudiu F. Bisu[a], Alain Gérard[a*], Jean-Yves K'nevez[a]

[a] University Politehnica from Bucharest, 313 Splaiul Independentei, 060042 Bucharest, Romania
[b] Université de Bordeaux, 351 cours de la Libération, 33405 Talence-cedex, France

**Abstract**

In the workspace model of machining an experimental procedure is implemented to determine the elastic behaviour of the machining system. In an other work we have proposed a static characterization of the machining system. In this paper a dynamic characterization and vibration analysis have long been used for the detection and identification of machine tool condition. The natural frequencies of the lathe machining system (Ernault HN400 - France) according with three different situations with no cutting process were acquired. The system modal analysis is used to identify the natural frequencies. These ones and these obtained on the spindle numerical model by finite element method are compared. This research is validated by experimental tests being based to measures of the lathe machine tool frequencies domain. Main focus is to identify a procedure to obtain natural frequencies values for machine tool components in order to establish better conditions in the cutting process on the machine tool.

**Nomenclature**

| | |
|---|---|
| **BT** | Block Tool |
| **BW** | Block Work piece |
| [C] | Damping matrix |
| $D_1$ | Holding fixture diameter (mm) |
| $D_2$ | Work piece diameter (mm) |
| $f_{sampling}$ | Sampling frequency (Hz) |
| $f_{max}$ | Highest frequency component in the measured signal (Hz) |
| [K] | Stiffness matrix (N/m) |
| $L_1$ | Holding fixture length (mm) |
| $L_2$ | Work piece length (mm) |
| [M] | Mass matrix (kg) |
| T | Period (s) |
| x (z) | Cross (feed) direction |
| y | Cutting axis |
| **WTM** | Work piece-Tool-Machine |
| $\omega_d$ | Damped natural frequency |
| $\xi_i$ | Percentage of damping |

*Corresponding author : E-mail: alain.gerard@u-bordeaux1.fr

# 1. Introduction

Among the various manufacturing processes to remove excess material, milling, drilling, grinding, turning occupies a choice place. This process is usually used in many circumstances as the outline of the parts for example, but also as soon as appear requirements of dimensional tolerance, precision or the surfaces quality of the produced part. However these dimensional accuracies or the final shape of the part (controlled roughness) is often dependent on the vibrations appearance during the process.

The vibration in the cutting process is a phenomenon dynamically unstable. These instabilities, often regenerative, are generated by many factors such as the work piece flexibility and the properties materials of the tools, the machine rigidity, tool geometry (approaching angle, rake angle, etc.) and the cutting tool edge radius, the nominal cutting conditions such as tool wear, feed rate and depth of cut, etc.

The cutting process stability is very studied for many reason of which, particularly, its the influence on the final surface quality (Chiou, et al. 1995; Karabay, 2007). The ordering of the process under various work conditions is an important problem for machining. Many modelings are implemented to optimise the cutting conditions, (Benardos, et al., 2006), (Karube, et al, 2002; Toh, 2004). The case of the orthogonal cut is the process that holds more the attention because of its simplicity to implement. For example by taking into account the regenerative aspect induced by the surface previously machined which has a sinusoidal form (waviness on the surface). The introduction of the non-linear interaction between the tool and carries it part always gives interesting results which go from regular (periodic or quasi-periodic) vibrations to possibilities of chaotic ones (Litak, 2002) or, always within the orthogonal framework of the cut, while utilizing in more the dry friction (Litak, et al., 2007).

The case 2D is also examined while considering carries it rigid part but by taking into account the tool flexibility (Insperger, et al., 2008), the tool holder flexibility (Chen et al., 2006) or the rotor system (Qi, et al., 200). Dassanayake et al. (2008) approach the case of the work piece dynamic response to the tool request that follows a regenerative surface. All these modelings give interesting results, but, in general, do not take account the whole of the kinematic chain of the cut process. We want to describe the whole of the kinematic chain of the cut process thereafter in a digital model richer and close to the experimental data.

Analysis of machines dynamic behaviour and equipments it's an important method to redesign the product or the manufacturing process and to assure the proper quality, maintenance and service. When we study machines or only part of them, dynamic behaviour is analysed for the following situations:

- Constant operating speed (e.g. speed rate of a rotor);
- Variable speed into a limited operating domain;
- Imposed speed inside the operational domain (e.g. rotational speed of the spindle 100 to 60,000 rpm);

In all of these mentioned cases the system behaviour under the external excitation effect is evaluated (Figure 1).

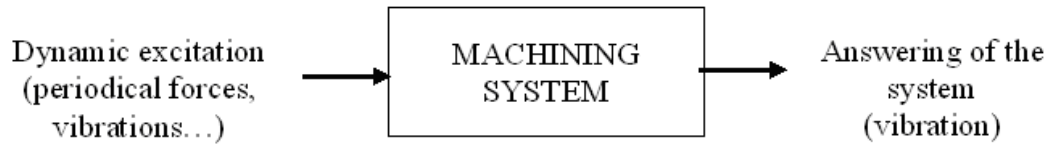

Figure 1 Transfer function

The transfer function is evaluated like the ratio response of the system / dynamic excitation. To diagnose one machine or equipment the main characteristics offered by transfer function are:

- Dynamic rigidity or compliance;
- Resonance frequencies;
- Damping factor;
- Natural modes of vibration.

The measurement of vibratory severity makes it possible to know if the vibratory behaviour of a machine exceeds the acceptable limits (Chen, et al., 2006). But within results sight, it is not possible to make an assumption on the vibration causes. This information could not be obtained that by using a frequency spectrum analysis (Moraru, et al., 1979).

We wish to be able to establish a model of three-dimensional turning processes nearest possible to the physical reality. Thus, the aim of this paper is to test a methodology that makes it possible to characterize the band with frequencies associated with the unit with each element of the kinematic chain apart from any operation of machining. Also, the tasks for performing the frequency analysis are given in Section 2. In Section 3 we present the experimental device with the frequency spectrum schema acquisition and the experimental results. A numerical model in finite element is used in order to find the first twelve natural frequencies. These ones confirm the first natural frequencies of the system spindle and the system spindle with work piece. The machine tool dynamic characterization is the object of Section 4 before to conclude.

## 2. Tasks for performing the frequency analysis

2.1 Technical determination of the frequency spectrum

The individual tasks for performing the frequency analysis are shown in the Figure 2.

Technical determination of the frequency spectrum is done nowadays with vibration sensors and FFT analysers. These modern analysers (LMS, Bruel & Kjaer, Schenck etc.) record the time signal of the vibration mixture in digital form, calculate the individual components and display the results in a frequency spectrum form.

All analysers and measurements can be triggered for start/stop, e.g., by RPM signal or vibration signal. Using a trigger, unwanted non-periodic components in the time signal and related spectrum, can be suppressed, e.g., for separation of other vibration signal.

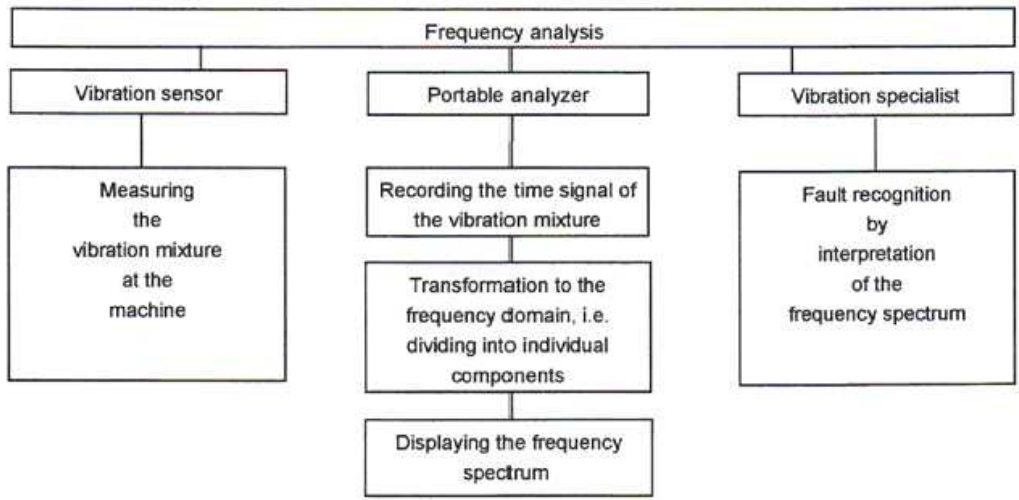

Figure 2 Frequency analysis tasks for a machine tool

2.2 Digitisation of the measured signals

The analogue signal from the vibration sensor is sent to the digital-to-digital converter (ADC) via an input amplifier, an anti-alias filter and a sample-and-hold circuit. This samples the signal and forms a time-sequential individual (discrete) amplitude value. This way each discrete value is quantified and converted to a binary data value (coding).

The *Shannon* theorem describes the important requirement for the sampling of analogue signals (Figure 3), namely regarding the relationship between the maximum signal frequency and the minimum sampling frequency. To guarantee the digitised signal reproducibility, the sampling frequency ($f_{sampling}$) must be at least double that of the highest frequency component ($f_{max}$) contained in the measured signal.

$$f_{sampling} \geq 2\, f_{max} \tag{1}$$

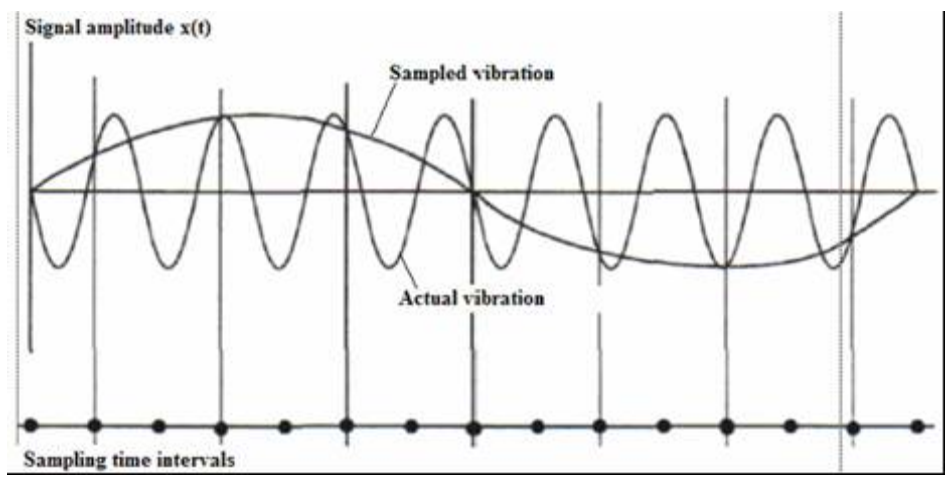

Figure 3 Example of signal vibration sampling

To counteract this undesirable effect (alias), all signal components with a frequency higher than half the sampling frequency must be suppressed. This is done in practice with a low-pass filter called an "anti-aliasing filter"

## 3. Experimental vibration analysis

3.1 Experimental device

All the machines vibrate and, as the state of the machines worsens (imbalance of the spindle or other important shaft, defect of bearing or spindle) the vibration level increases. While measuring and by supervising the vibration level produced by a machine, is obtained an ideal indicator on his state, especially dynamic behaviour (Karabay, 2007; Cardi, 2008).

While the increase in machine vibration makes it possible to detect a defect, the analysis of the machine vibration characteristics makes it possible to identify the cause of it. In this way we can deduce with sufficient precision the time domain before vibrations does not become critical.

Each component of spectrum FFT corresponds to a characteristic frequency well defined (imbalance, resonance, misalignment etc.) (Lalwani, et al., 2008). The analysis in frequency is carried out in general when the machine vibratory level is considered to be higher than the acceptable threshold (Rigal, et al., 1998).

In the Figure 4 is illustrated the schema acquisition of the frequency spectrum FFT in the case of the Ernault HN400 lathe - France, LMP Laboratory from Bordeaux 1 University. Accelerometers were fixed on the horizontal and vertical plans on the bearing fix front part of the lathe spindle.

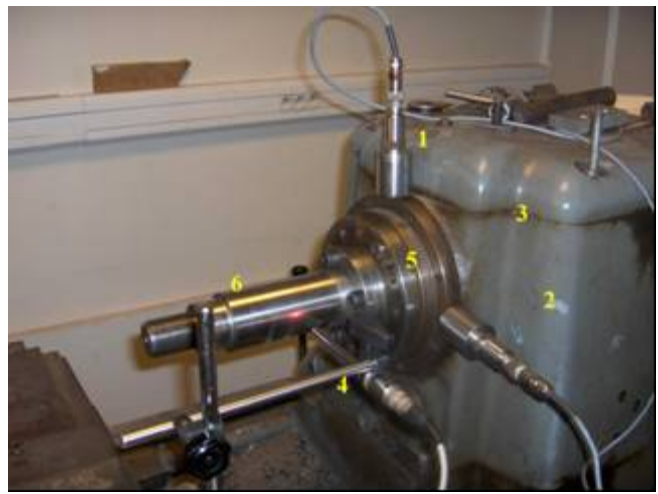

Figure 4 Position of transducers in acquisition schema to acquire vibrations of the spindle ERNAULT. 1- Accelerometer AS020 -vertical plan (V); 2- Accelerometer AS020 -horizontal plan (H), 3- Gearbox of Ernault HN400 lathe; 4- Tachometer; 5- Spindle bearing front side; 6- Work piece.

An example of the vibration level of this spindle in horizontal plan is illustrated in the Figure 5. Tracking vibration signal versus time for different spindle speed levels was used. The vibration speed level in mm/s (rms) is situated on the range 0,1-0,3. For this type of machine tool, in accordance with IEC 34-14 standard the upper limit of the vibration level is 1,8 mm/s (see Table 1).

| Related operating speed (rpm) | Limit value of the effective vibration velocity for shaft heights H in mm | | | |
|---|---|---|---|---|
| Normal | Machine is mounted on specifically design elastic foundations (e.g. vibration dampers) | | | Specifically designed rigid |
| (mm/s) | 56 < H < 132 | 132 < H < 225 | H > 225 | H > 40 |
| > 600 < 1.800 | 1,8 | 1,8 | 2,8 | 2,8 |
| > 1.800 < 3.600 | 1,8 | 2,8 | 4,5 | 2,8 |

Table 1 Limit values in accordance with IEC 34-14.

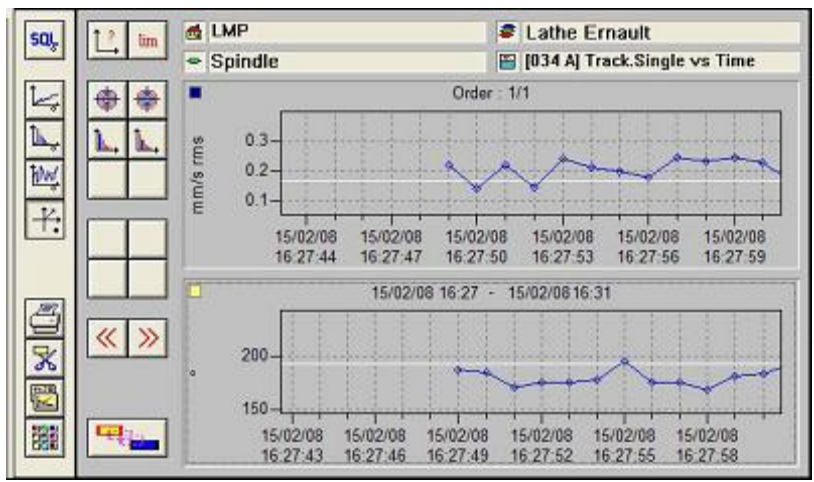

Figure 5 Vibration level of the Ernault lathe using Tracking versus time procedure

3.2 Block Work piece: **BW**

As many workers (Benardos, et al., 2006), (Yaldiz, et al. 2006), (Mehdi, et al., 2002a) a cylindrical geometry of the work piece is chosen. The Block Work piece (**BW**) represents the revolving part of the Machine-Tool-Machine system (**WTM**); it includes the holding fixture, the work piece and the spindle Figure 6. To make the whole frame rigid, a very rigid unit (work piece, holding fixture) is conceived in front of the **WTM** elements (Figure 7).

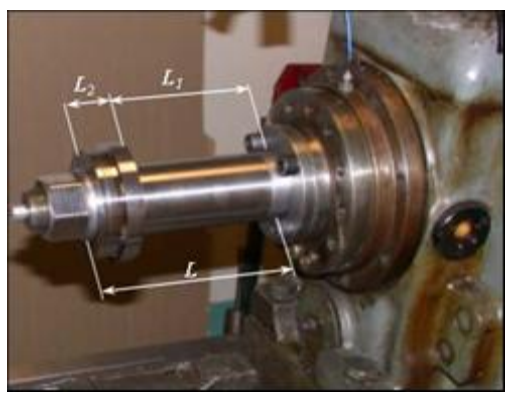

Figure 6 Representation of the **BW**

The work piece geometry and his holding fixture are selected with $D_1 = 60$ mm, $D_2 = 120$ mm and $L_2 = 30$ mm (cf. Figure 7). These dimensions retained for these test tubes were selected using the finite elements method coupled to an optimisation method by SAMCEF software. It is necessary to determine the holding fixture length $L_1$ to obtain a significant stiffness in flexion. The objective is to move away the first vibration mode of the Block work piece of the lathe vibration fundamental natural mode (see Bisu, 2007).

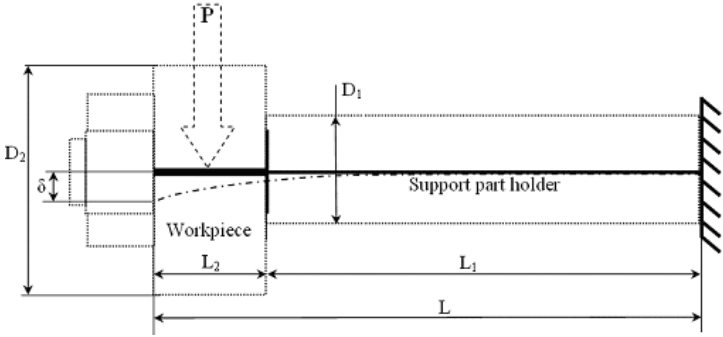

Figure 7 Geometry of holding fixture / work piece

The stiffness is calculated on the basis of the displacement δ for a given force P :

$$\delta = \frac{P \times L^3}{3E \times I} \qquad (2)$$

with inertial moment :

$$I = \frac{\pi \times D_1^4}{64} \qquad (3)$$

The Figure 8 represents the displacements and the stiffness values relating to the length of holding fixture / work piece, for a force $P = 1,000$ N, a Young modulus $E = 21.10^5$ N/mm² and a holding fixture diameter $D_1 = 60$ mm.

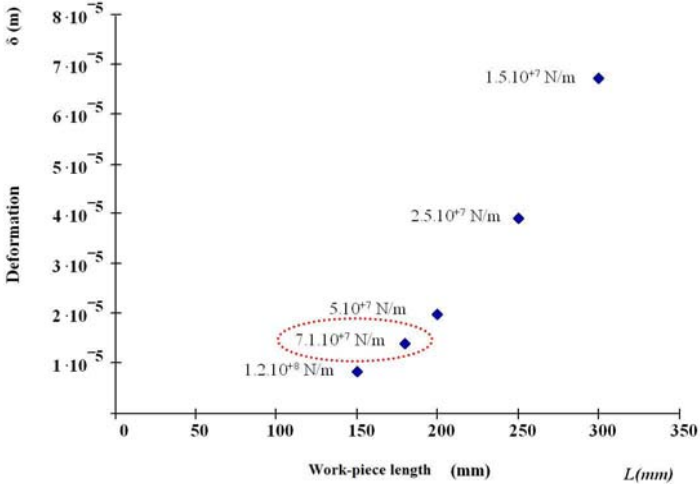

Figure 8 Displacements according to the holding fixture length

A holding fixture length: $L_1 = 180$ mm, for a stiffness in flexion of $7.10^7$ N/m, is reminded. This value is in the higher limit of the acceptable zone of rigidity for conventional lathe (cf. Figure 9), (Ispas, et al., 1999; Koenigsberger, et al., 1970; Konig, et al., 1997).

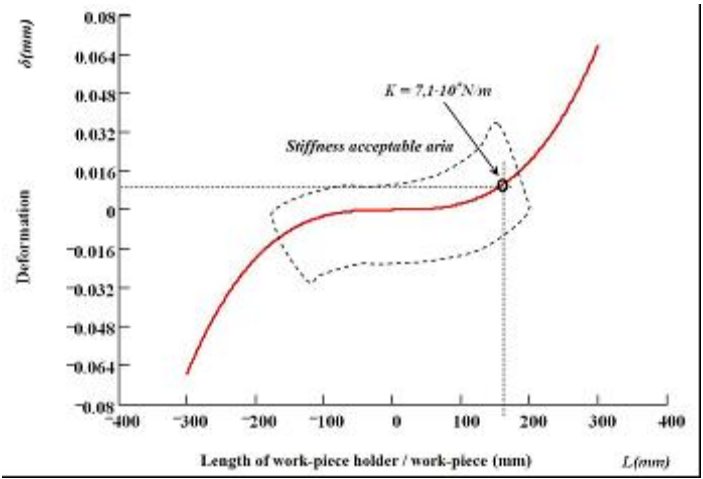

Figure 9 Representation of the acceptable aria of the work piece deformation

In this case, the **BT** part includes the tool, the tool-holder, the dynamometer, the fixing plate on the cross slide (Figure10). The six-components dynamometer (Couetard-00) is fixed between the cross slide and the tool-holder. It is necessary thereafter for the measurement of the mechanical actions of cutting.

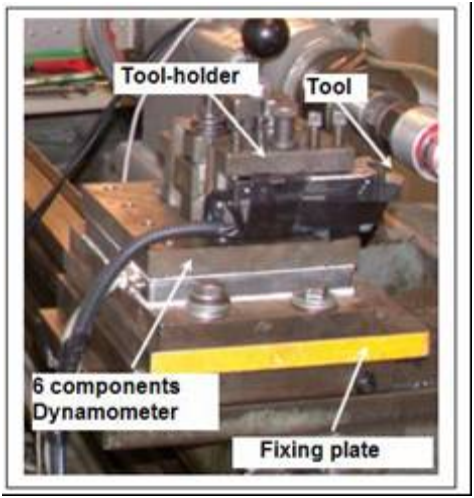

Figure 10 Representation of the block tool **BT**

3.4 Experimental results

The natural frequencies experimental values obtained following the impact tests are detailed in the Figure 11 and Figure 12, considering blocks **BT** and respectively **BW**. Using an impact hammer, the natural frequencies of each block, for each element, are given accordingly the machine axes directions x (cross direction), y (cutting direction), and z (feed direction). A three directions accelerometer is positioned on each element, in each direction and the elements tested were the subject of the hammer impact. An example of result is given into the Figure 11, where the natural frequencies of whole block **BT** are presented, with all the components.

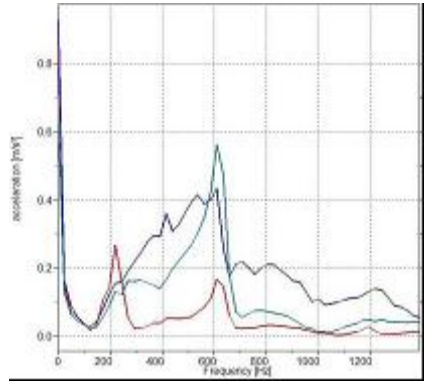

Figure 11 Natural frequencies representation for the block-tool **BT** (direction: x red colour, y blue colour, z green colour)

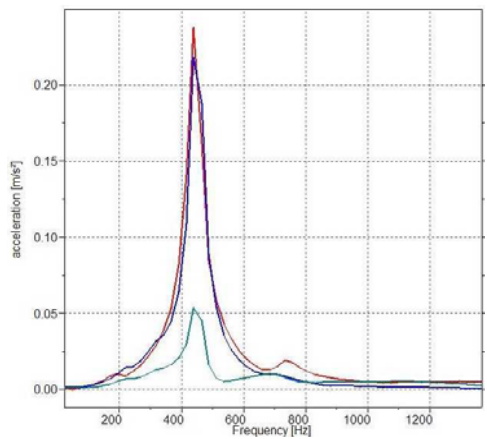

Figure 12 Natural frequencies for the block-work-piece **BW** ((direction: x red colour, y blue colour, z green colour)

For the **BW** unit the results are presented in the Figure 12. The natural frequencies domain is presented for each component of the system by carrying out a modal superposition in the Figure 13. These results are coherent with those met in the literature (Moraru, et al., 1979; Ispas, et al., 1999; Marinescu, et al., 2002; Benmohammed, 1996; Kudinov, 1970).

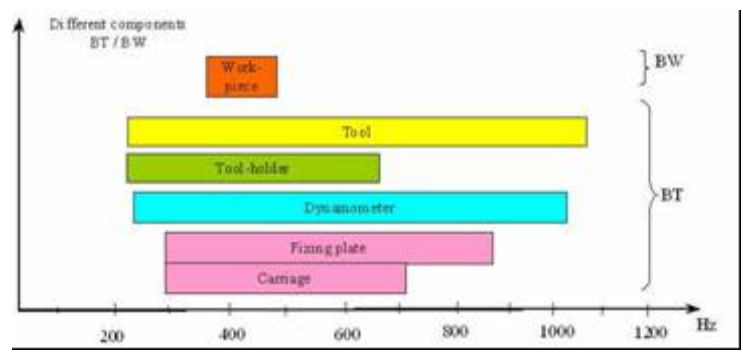

Figure 13 Natural frequencies domains superposition of the machine tool components at the time of the impact for each element.

Using the layout of the tool free oscillations, (presented in the Figure 14), the $\xi_i$ damping percentage is directly given by the logarithmic decrement curve, starting from *N* consecutive maximum; the measured period $T$ of the damped oscillations give the damped natural frequency $\omega_d$ of the system.

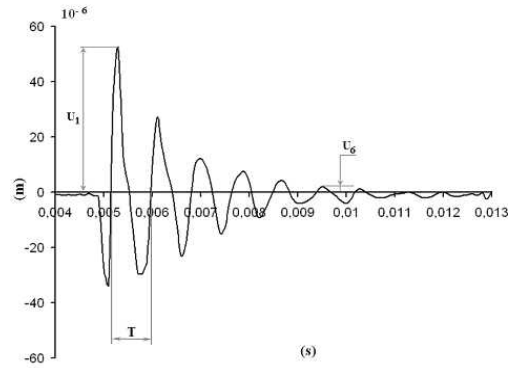

Figure 14 Test of free displacement during of one impulse using an impact hammer.

With the obtained values $\omega_d$ and $\xi_i$, we can calculate no damped natural frequency. The values of these parameters allow calculating: the stiffness $K$, the equivalent mass $M$ and the equivalent damping coefficient $C$ for each part **BT** and **BW** and for all three directions. Like values example it comes:

$$M = \begin{bmatrix} 2.2 & 0 & 0 \\ 0 & 5.3 & 0 \\ 0 & 0 & 2.5 \end{bmatrix} \quad (4)$$

$$C = \begin{bmatrix} 1.2 \times 10^3 & 0 & 0 \\ 0 & 0.89 \times 10^3 & 0 \\ 0 & 0 & 2.5 \times 10^3 \end{bmatrix} \quad (5)$$

and the stiffness matrix $k$ is given in the first part of this work (Bisu, et al., 2007).

3.5 Numerical model of the assembly of the lathe spindle Ernault HN400.

In order to find the spindle natural frequencies domain of the machine tool and of the assembly Block Work piece **BW** (spindle with the work-piece attached) was used Finite Element Method. The Figure 15 shows the mesh (4,290 nods, 16,310 nets) and spindle numerical model. The Figure 16 shows the model by adding the work-piece.

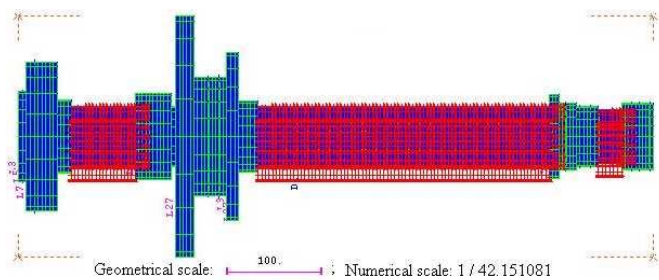

Figure. 15 Spindle numerical model.

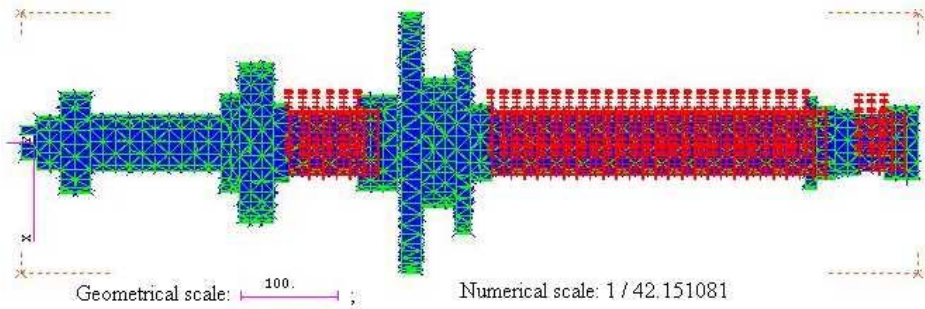

Figure 16 Spindle with work piece **BW** numerical model and boundary conditions.

Accordingly numerical models were obtained the results for the first twelve natural frequencies of these two models (represented in the Figure 15 and Figure 16). Vibration mode accordingly first natural frequency is represented in the Figure 17 and the numerical values of first twelve natural frequencies are represented in the Figure 18.

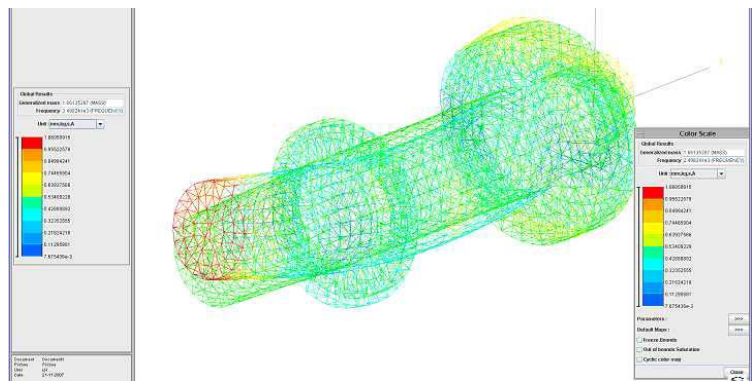

Figure 17 Vibration bending mode for the first natural frequency

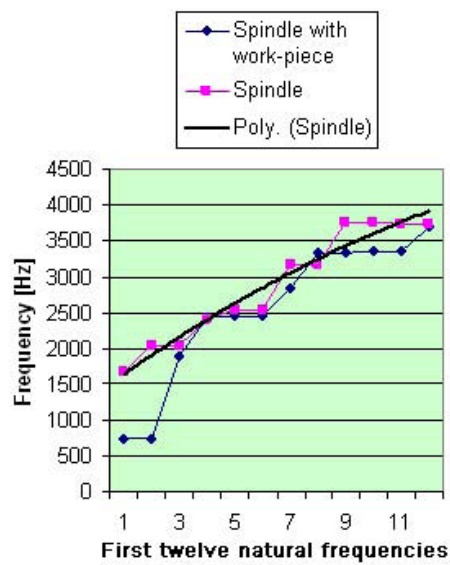

Figure 18 Spindle natural frequencies polynomial trend of the machine tool

We can observe that the first spindle natural frequency of the system with work piece is about 675 Hz and it is lower considering only the spindle system (1,600 Hz). These results are coherent with those experimentally obtained (Figure 19).

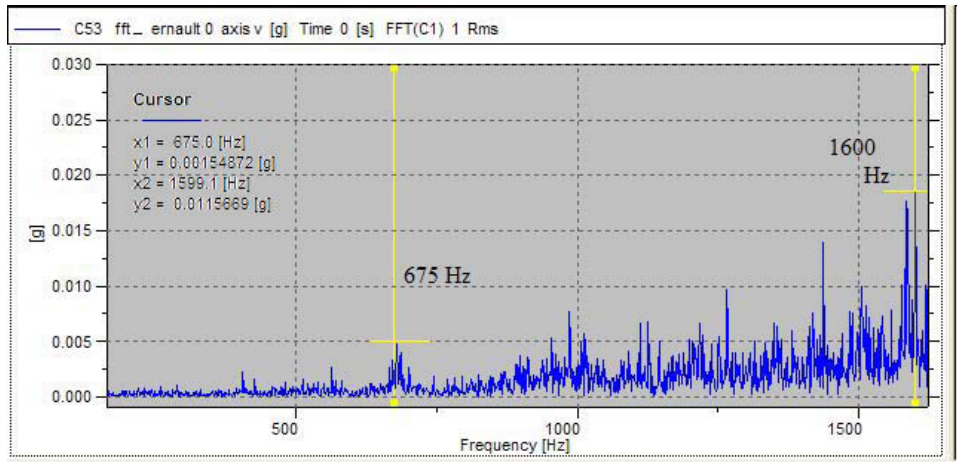

Figure 19 Experimentally FFT spectrum; spindle turning at 1,000 rpm

On the acquired signal FFT spectrum (spindle system turning, without cutting process) we can see the first frequency 675 Hz and, on the right side, the frequency with the value 1,599.1 Hz (corresponding to the spindle alone).

## 4. Machine tool dynamic characterization - turning process recommendations

4.1 An analysis according to three configurations

The dynamic characterization of the machining system is supplemented by an analysis according to three configurations: electric motor drive, electric motor turning the spindle, electric motor turning the spindle and advance movement coupled (Bisu, 2007, Bisu, et al., 2006). The three axes accelerometer is placed on the tool body and the one-way accelerometer is located on the bed near the spindle. In the Figure 20 are presented the frequencies values in the case when the machine tool is no charge with the cutting process, for the three configurations mentioned. The frequencies below 100 Hz, belong to the engine behaviour, the amplitudes are very weak, very low and appear on each tests configuration.

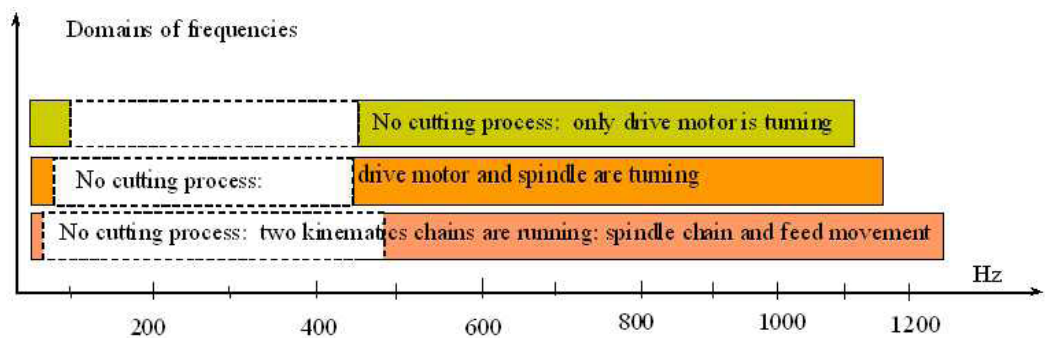

Figure 20 Frequencies domains representation in the case with no cutting process

The frequencies corresponding with no cutting process are given for the three directions in the case of the kinematics chain "electric motor turning the spindle and advance movement coupled" (Bisu, 2007). The measured significant frequencies are into the domain of 230 Hz up to 1,000 Hz (see Figure 21), and more, on the three x, y and z directions.

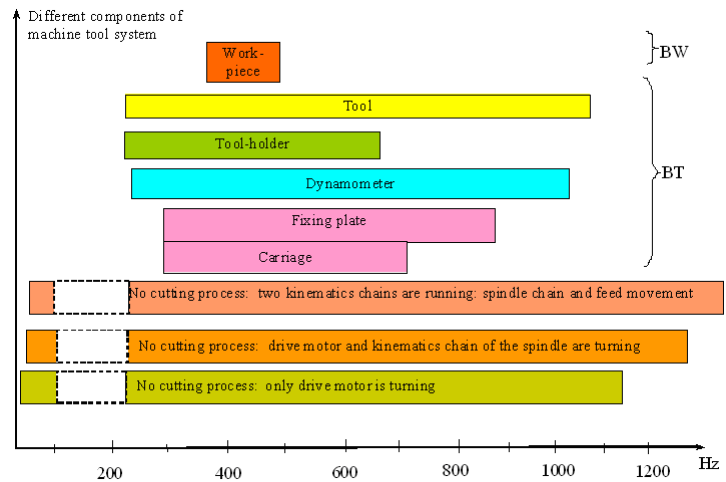

Figure 21 Natural frequencies superposition in the case of no cutting process

4.2 About a better manage the cutting process

In order to have got a better manage the cutting process, next recommendations are considered:

*Increasing the work piece rigidity.* In order to reduce the vibrations problems at the time of the machining of the thin walls it is always well adapted, like known as previously, to rigidify the part by an adapted fixture, when that is possible (Mehdi, 2002a; Sutter, et al., 2005; Mehdi, et al., 2002b). The process simulations makes possible to specify before to have begun machining the most adapted solution, and even it is possible to redesign the shape of the work-piece in order to bring more rigidity to him.

*Control of the cutting stiffness* by reducing axial tool engagement. The theory of the stability lobes highlighted very early the importance of the relationship between the work-piece stiffness or the tool, and the cutting stiffness (the coefficient which links the tool displacement into the material to the cutting pressure, in the vibration direction). The most significant parameter to decrease this ratio is roughly speaking the projected cutting tool edge length in the material, which always impose to the specialists in machining process to naturally reducing the tool axial engagement (Calderon, 1998).

*Modifying the cutting tool angles.* Another way of decreasing the cutting stiffness is to exploit the tool angles so that the cutting pressure is parallel with the wall, thus the cutting stiffness would be theoretically null.

These two possibilities of stiffness control are major in the vibratory phenomenon during machining, in tools selection (angles, coatings etc.) and the strategies of machining in generally have the tendency to exploit these two effects (Belhadi, et al., 2005), (Mabrouki, et al., 2004}.

**Limit the possibility to generate vibrations, avoid the resonance.**

- Avoiding the frequencies with problems: by choosing the proper speed of the spindle (considering the stability lobes theory);

- Adding damping: by lubrication, controlling the pressure on the assembly surfaces, etc;
- Obtaining the excitation spectrum: by tools with variable step, by the spindle rotation with variable speed;
- Controlling actively: by the use of actuators (axes machine, turret etc.) controlled dynamically according to measurements, in order to eliminate the vibrations. For the moment, in practice, the achievements in this field are mainly regulations of average effort or power by the reduction of the tool advance (Karabay, 2007, Bakkal et al., 2004.

Within the framework of this study, the objective was to find (starting from the experimental tests in turning) a correlation between the frequency founded on the acquired FFT spectrum and the dynamic process based on the elastic system Machine tool, Work piece and Tool device.

Machine tools and particularly, in this work paper, the lathe, evolved to the limits of certain parameters of cutting that consist in a better control of the process. In this context, this subject "dynamic characterization" is very important and could helping on the development of a simulation model for the cutting process on the machine tool.

This research was validated by experimental dynamic tests being based to measures of the frequencies domain of the machining system Ernault HN400 lathe, in the LMP laboratory of the University Bordeaux 1, France. Nevertheless the method used here applies to other machine tools (milling, drilling etc.).

Taking into account the recommendations summarized in the Section 4 the cutting process could be better managed.

The following step in our research will be to make a prediction for the cutting conditions and to supplement necessary knowledge toward a global and industrial model of cutting process considering a 3D model in turning.

## Acknowledgements

The authors acknowledge Jean Pierre Larivière Engineer Centre National de la Recherche Scientifique - France for the numerical simulation with SAMCEF software. The authors would like also to thank the CNRS (UMR 5469) for the financial support to accomplish the project.